\documentclass[entropy,article,accept,oneauthor,pdftex,10pt,a4paper]{mdpi}
\lastpage{}
\doinum{ To be determined}       
\pubvolume{}    
\pubyear{2017}
\history{Accepted 08/2017}

\usepackage{dsfont}
\usepackage[utf8]{inputenc}
\usepackage{algorithm, mathtools}
\usepackage[noend]{algpseudocode}
\usepackage{amsmath}
\usepackage{amssymb}
\usepackage{natbib}
\usepackage{enumerate}
\usepackage{microtype}

\DeclareUnicodeCharacter{00A0}{ }

\definecolor{orange}{rgb}{1, 0.5, 0}
\definecolor{green}{rgb}{0, 0.5, 0}
\definecolor{darkblue}{rgb}{0, 0, 0.7}

\newcommand{\revision}{}
\renewcommand{\d}{\boldsymbol{d}}

\newcommand{\x}{\boldsymbol{\theta}}
\newcommand{\bphi}{\boldsymbol{\phi}}
\newcommand{\boldeta}{\boldsymbol{\eta}}

\newcommand{\depth}{(\textnormal{depth})}
\newcommand{\xref}{\x_{\rm ref}}
\newcommand{\ytot}{y_{\rm tot}}
\newcommand{\ztot}{z_{\rm tot}}

\newcommand{\apj}{The Astrophysical Journal}
\newcommand{\mnras}{Monthly Notices of the Royal Astronomical Society}

\Title{Computing Entropies With Nested Sampling}

\Author{Brendon J. Brewer}

\address{Department of Statistics, The University of Auckland, Private Bag 92019,
Auckland 1142, New Zealand}

\corres{{\tt bj.brewer@auckland.ac.nz}}

\abstract{The Shannon entropy, and related quantities such as mutual
information, can be used to quantify uncertainty and relevance.
However, in practice, it can be difficult to compute these quantities
for arbitrary probability distributions, particularly if the probability
mass functions or densities cannot be evaluated. This paper introduces a
computational approach, based on Nested Sampling, to evaluate entropies of
probability distributions that can only be sampled. I demonstrate the method
on three examples: a simple gaussian example where the key quantities are
available analytically; (ii)
an experimental design example about scheduling observations in order
to measure the period of an oscillating signal; and (iii) predicting the
future from the past in a heavy-tailed scenario.}


\begin{document}


\section{Introduction}

If an unknown quantity $x$ has a discrete probability distribution $p(x)$,
the Shannon entropy \citep{shannon, cover2012elements} is defined as
\begin{align}
H(x) &= -\sum_{x} p(x) \log p(x)
\end{align}
where the sum is over all of the possible values of $x$ under consideration.
The entropy quantifies the degree to which the issue
``what is the value of $x$, precisely?'' remains unresolved
\citep{knuth_questions}. {\revision More generally, the `Kullback-Leibler divergence'
quantifies the degree to which a probability distribution $p(x)$ departs
from a `base' or `prior' probability distribution $q(x)$:
\begin{align}
D_{\rm KL}\big(p \,||\, q\big) &=
    \sum_{x} p(x) \log\left[\frac{p(x)}{q(x)}\right]
\end{align}
\citep{knuth2012foundations, caticha2006updating}.
If $q(x)$ is taken to be uniform, this is equivalent (up to an additive
constant and a change of sign) to the Shannon entropy $H(x)$.
With the negative sign reinstated, Kullback-Leibler divergence is sometimes
called relative entropy, and is more fundamental than the Shannon entropy
\citep{knuth2012foundations}. However, this paper focuses on the Shannon
entropy for simplicity, and because the resulting algorithm calculates
the expected value of a log-probability.

The Shannon entropy can be interpreted straightforwardly as a measure of
uncertainty.} For example,
if there is only one possible value which has probability one, $H(x)$ is
zero. Conventionally, $0 \times \log 0$ is defined to be equal
to $\lim_{x \to 0^+} (x\log x)$, which is zero; i.e., `possibilities' with
probability zero do not contribute to the sum.
If there are $N$ possibilities with equal probabilities $1/N$ each,
then the entropy is $H(x) = \log(N)$.
If the space of possible $x$-values is continuous so that $p(x)$ is a
probability density function, the differential entropy
\begin{align}
H(x) &= -\int p(x) \log p(x) \, dx
\end{align}
quantifies uncertainty by generalising the log-volume
of the plausible region, where volume is defined with respect to $dx$.
{\revision Crucially, in continuous spaces, the differential entropy is not
invariant under changes of coordinates, so any value given for a
differential entropy must be understood with respect to the coordinate
system that was used. Even in discrete cases, using the Shannon entropy assumes
that each possibility contributes equally when counting possibilities ---
if this is inappropriate, the relative entropy should be used.}

Entropies, {\revision relative entropies}, and related quantities,
tend to be analytically available only for
a few families of probability distributions. On the numerical side, if
$\log p(x)$ can be evaluated for any $x$, then simple Monte Carlo will suffice
for approximating $H(x)$. On the other hand, if $p(x)$ can only be
{\em sampled} but not evaluated (for example, if $p(x)$ is a marginal {\revision or
conditional distribution, or the distribution of a quantity derived from $x$)},
then this will not work. Kernel density estimation or the like
\citep[e.g.][]{JMLR:v15:szabo14a}
may be effective in this case, but is unlikely to generalise well to high
dimensions.

This paper introduces a computational approach to evaluating the Shannon
entropy $H(x)$ of probability distributions that can only be sampled
using Markov Chain Monte Carlo (MCMC). {\revision The key idea is that Nested Sampling
can create an unbiased estimate of a {\em log}-probability, and entropies
are averages of log-probabilities}.

\subsection{Notation and conventions}

Throughout this paper, I use the compact
`overloaded' notation for probability distributions favoured by many
Bayesian writers \citep{jaynes2003probability, mackay2003information},
writing $p(x)$ for either a probability mass function
or a probability density function, instead
of the more cumbersome $P(X=x)$ or $f_X(x)$.
In the compact notation, there is no distinction between the
`random variable' itself ($X$) and a similarly-named dummy variable ($x$).
Probability distributions are implicitly conditional on some prior
information, which is omitted from the notation unless necessary.
All logarithms are written $\log$ and any base can be used, unless otherwise
specified (for example by writing $\ln$ or $\log_{10}$). Any numerical
values given for the value of specific entropies are in {\em nats}, i.e.,
the natural logarithm was used.

Even though the entropy is written as $H(x)$, it is imperative that we
remember it is not a property of the value of $x$ itself, but a property
of the probability distribution used to describe a state of ignorance about $x$.
Throughout this paper, $H(x)$ is used as notation for both Shannon entropies
(in discrete cases) and differential entropies (in continuous cases). Which
one it is should be clear from the context of the problem at hand.
I sometimes write general formulae in terms of sums (i.e., as they would
appear in a discrete problem), and sometimes as integrals, as they would
appear in a continuous problem.

\section{Entropies in Bayesian inference}

Bayesian inference is the use of probability theory to
describe uncertainty, often about unknown quantities
(`parameters') $\x$. Some data $\d$, initially unknown but
thought to be relevant to $\x$, is obtained.
The prior information leads the
user to specify a prior distribution $p(\x)$ for the unknown parameters,
along with a conditional distribution $p(\d | \x)$ describing prior knowledge
of how the data is related to the parameters
(i.e., if the parameters were known, what data would likely be observed?).
By the product rule, this yields a {\em joint prior}
\begin{align}
p(\x, \d) &= p(\x)p(\d | \x)
\end{align}
which is the starting point for
Bayesian inference \citep{caticha2008lectures, caticha2006updating}.
In practical applications, the following operations
are usually feasible and have a low computational cost:
\begin{enumerate}
  \item {\revision Samples} can be generated from the prior $p(\x)$;
  \item Simulated datasets can be generated from $p(\d | \x)$ for any
        given value of $\x$;
  \item The likelihood, $p(\d | \x)$, can be evaluated cheaply for any
        $\d$ and $\x$. Usually it is the log-likelihood that is actually
        implemented, for numerical reasons.
\end{enumerate}
Throughout this paper I assume these operations are available and inexpensive.

\subsection{The relevance of data}

The entropy of the prior $p(\x)$ describes the degree to which the question
``what is the value of $\x$, precisely?'' remains unanswered, while the
entropy of the joint prior $p(\x, \d)$
describes the degree to which the question
``what is the value of the pair $(\x, \d)$, precisely?'' remains unanswered.
The degree to which the question ``what is the value of $\x$?'' would remain
unresolved if $\d$ were resolved is given by the
conditional entropy
\begin{align}
H(\x | \d) &= - \sum_{\d} p(\d) \sum_{\x} p(\x | \d) \log p(\x | \d)
\end{align}
which is the expected value of the entropy of the posterior, averaged over
all possible datasets which might be observed. 

One might wish to compute $H(\x | \d)$, perhaps to compare it to
$H(\x)$ and quantify how much might be learned about $\x$.
This would be difficult because the expression
for the posterior distribution
\begin{align}
p(\x | \d) &= \frac{p(\x)p(\d | \x)}{p(\d)}
\end{align}
contains the marginal likelihood:
\begin{align}
p(\d) &= \sum_{\x} p(\x) p(\d | \x)
\end{align}
also known as the `evidence', which tends to be computable but costly,
especially when the summation is replaced by an integration over a
high-dimensional parameter space.

It is important to distinguish between $H(\x | \d)$ and the
entropy of $\x$ given a particular value of $\d$, which might be written
$H(\x | \d=\d_{\rm obs})$ (`obs' for observed).
The former measures the degree to which one question answers
another {\em ex ante}, and is a function of two questions.
The latter measures the degree to which {\revision a question remains unresolved
after conditioning on a specific statement ({\em ex post})},
and is a function of a question and a statement.

\subsection{Mutual information}

The mutual information is another way of describing the relevance of the
data to the parameters. Its definition, and relation to other quantities, is
\begin{align}
I(\x; \d) &= \sum_{\x} \sum_{\d} p(\x, \d)
                       \log\left[\frac{p(\x, \d)}{p(\x)p(\d)}\right]\\
           &= H(\x) + H(\d) - H(\x, \d)\\
           &= H(\d) - H(\d | \x)\\
           &= H(\x) - H(\x | \d).
\end{align}
The mutual information can also be written as the expected value
(with respect to the prior over datasets $p(\d)$) of the Kullback-Leibler
divergence from prior to posterior:
\begin{align}
I(\x ; \d) &= \sum_{\d} p(\d) D_{\rm KL}\big(p(\x|\d) \, || \, p(\x)\big).
\end{align}
In terms of the prior, likelihood, and evidence,
it is
\begin{align}
I(\x; \d) &= \sum_{\x} \sum_{\d} p(\x)p(\d | \x)
              \left[\log p(\d | \x) - \log p(\d)\right],
\end{align}
i.e., the mutual information is the prior expected value of the
log likelihood minus the log evidence.
As with the conditional entropy, the computational
difficulty appears in the form of the log evidence,
$\log p(\d) = \log \left(\sum_{\x} p(\x)p(\d | \x)\right)$, which must be evaluated
or estimated for many possible datasets.

For experimental design purposes, maximising the expected amount
of information obtained from the data is a sensible goal.
Formally, either maximising $I(\x; \d)$ or
minimising $H(\x | \d)$ will produce the same result because the prior $p(\x)$
does not
vary with the experimental design. Reference priors
\citep{bernardo2005reference} also maximise $I(\x; \d)$ but vary the prior
$p(\x)$
in the maximisation process while keeping the experimental design fixed.

\section{Nested Sampling}

Nested Sampling (NS), introduced by
\citet{skilling2006nested}, is an algorithm whose aim is
to calculate the evidence {\revision
\begin{align}
p(\x) &= \int p(\x) p(\d | \x) \, d\x,
\end{align}
or, in the simplified notation common when discussing computational matters,
\begin{align}
Z &= \int \pi(\x) L(\x) \, d\x,
\end{align}}
\noindent where $\x$ is the unknown parameter(s), $\pi$ is the prior distribution,
and $L$ is the likelihood function. The original NS algorithm, and variations
within the same family \citep{feroz2009multinest, dns, handley2015polychord},
have become popular tools in Bayesian data analysis
\citep{knuth2015bayesian, pullen2014bayesian, exoplanet} and have also
found use in statistical physics \citep{partay2010efficient,
baldock2016determining, martiniani2014superposition}, which was another of
its aims.
{\revision
NS also estimates the Kullback-Leibler divergence from the prior
$\pi(\x)$ to the posterior $\pi(x)L(x)/Z$, which Skilling calls the
information. This is a measure of how much was learned about $\x$
from the specific dataset, and is also useful for defining a termination
criterion for NS.}

While $Z$ is a simple expectation with
respect to $\pi$, the
{\revision distribution of $L$-values implied by $\pi$}
tends to be very heavy-tailed, which is
why simple Monte Carlo does not work. Equivalently, the integral 
{\revision over $\x$} is dominated by very small regions where $L(\x)$ is high.

To overcome this, NS evolves a population of $N$ particles in the parameter
space towards higher likelihood regions.
The particles are initially initialised from the prior $\pi(\x)$, and the
particle with the lowest likelihood, $L^*_1$, is found and its details are
recorded as output. This worst particle is then discarded
and replaced by a new particle
drawn from the prior $\pi$ but subject to the constraint that its likelihood
must be above $L^*_1$. {\revision This is usually achieved by cloning a surviving particle
and evolving it with MCMC to explore the prior distribution but with
a constraint on likelihood value, $L(\x) > L^*_1$. This `constrained prior'
distribution has probability density proportional to
\begin{align}
\pi(\x)\mathds{1}\left(L(\x) > L^*_1\right)
\end{align}
where $\mathds{1}()$ is an indicator function which returns one if the argument
is true and zero if it is false.}

This process is repeated, resulting in an
increasing sequence of likelihoods
\begin{align}
L^*_1, L^*_2, L^*_3,... \label{eqn:likelihood_sequence}
\end{align}
Defining $X(\ell)$ as the amount of prior mass with likelihood greater than
some threshold $\ell$,
\begin{align}
X(\ell) &= \int \pi(\x) \mathds{1}\left(L(\x) > \ell\right) \, d\x,
\end{align}
each discarded point in the sequence can be assigned an $X$-value,
which transforms $Z$ to a one-dimensional integral.
\citet{skilling2006nested}'s key insight was that the
$X$-values of the discarded points are approximately known from the nature of the
algorithm. Specifically, each iteration reduces the prior mass by approximately
a factor $e^{-1/N}$. More accurately, the conditional distribution of
$X_{i}$ given $X_{i-1}$ (and the definition of the algorithm)
is obtainable by letting
$t_i \sim \textnormal{Beta}(N,1)$ and setting $X_{i+1} := t_iX_{i-1}$ (where
$X_1 = t_1$).
The $t_i$ quantities are the proportion of the remaining prior mass
that is {\em retained} at each iteration, after imposing the constraint of the
latest $L^*$ value.

\subsection{The sequence of $X$ values}

{\revision
The sequence of $X$ values is defined by
\begin{align}
X_1 &= t_1 \\
X_2 &= t_2X_1 \\
X_3 &= t_3X_2
\end{align}
and so on, i.e. $X_i$ can be written as
\begin{align}
X_i = \prod_{k=1}^i t_k.
\end{align}
Taking the log and multiplying both sides by -1, we get
\begin{align}
-\ln X_i = \sum_{k=1}^i \left(-\ln t_k\right).\label{eqn:minuslogX}
\end{align}

Consider $t_i$, the fraction of the remaining prior mass that is retained
in each NS iteration. By a change of variables,
the distribution for these compression factors $t_i \sim $Beta$(N,1)$
corresponds to an exponential distribution for $-\ln(t_i)$:
\begin{align}
-\ln t_i &\sim \textnormal{Exponential}(N).\label{eqn:exp}
\end{align}

From Equations~\ref{eqn:minuslogX} and~\ref{eqn:exp}, we see that}
the sequence of $-\ln X$ values produced by a Nested Sampling run can be
considered as a Poisson process with rate equal to $N$, the number of
particles. This is why separate NS runs can be simply merged
\citep{skilling2006nested, HENDERSON201784}.
\citet{Walter2017point} showed how this view of the NS sequence of points
can be used to construct a
version of NS that produces unbiased (in the frequentist sense)
estimates of the evidence.
However, it can also be used to construct an unbiased estimator of
a log-probability, which is more relevant to information theoretic
quantities discussed in the present paper.

Consider a particular likelihood value $\ell$ whose corresponding $X$-value
we would like to know. Since the $-\ln X_i$ values have the same distribution
as the arrival times of a Poisson process with rate $N$, the probability
distribution for the number of points in an interval of length $w$ is
Poisson with expected value $Nw$. Therefore the expected number of
NS discarded points with likelihood below $\ell$ is $-N\ln X(\ell)$:
\begin{align}
\big< n(L(\x_i) \leq \ell) \big> &= -N\ln X(\ell).
\end{align}
We can therefore take $n(L(\x_i) \leq \ell)/N$, the number of points in the
sequence with likelihood below $\ell$ divided by the number of NS particles,
as an unbiased estimator of $-\ln X(\ell)$. The standard deviation of the
estimator is $\sqrt{-\ln X(\ell)/N}$.

\section{The algorithm}

The above insight, that the number of NS discarded points with likelihood
below $\ell$ has expected value $-N\ln X(\ell)$, is the basis of the algorithm.
If we wish to measure a log-probability $-\ln X(\ell)$, we can use NS to do it.
$\pi(\x)$ and $L(\x)$ need not be the prior and likelihood respectively, but
can be any probability distribution (over any space) and any function whose
expected value is needed for a particular application.

See Algorithm~\ref{alg:algorithm} for a step-by-step description of the
algorithm. The algorithm is written in terms of an unknown quantity $\x$
whose entropy is required. In applications, $\x$ may be parameters,
a subset of the parameters, data, some function of the data,
parameters and data jointly, or whatever.

\begin{figure}[!ht]
\centering
\includegraphics[width=0.5\linewidth]{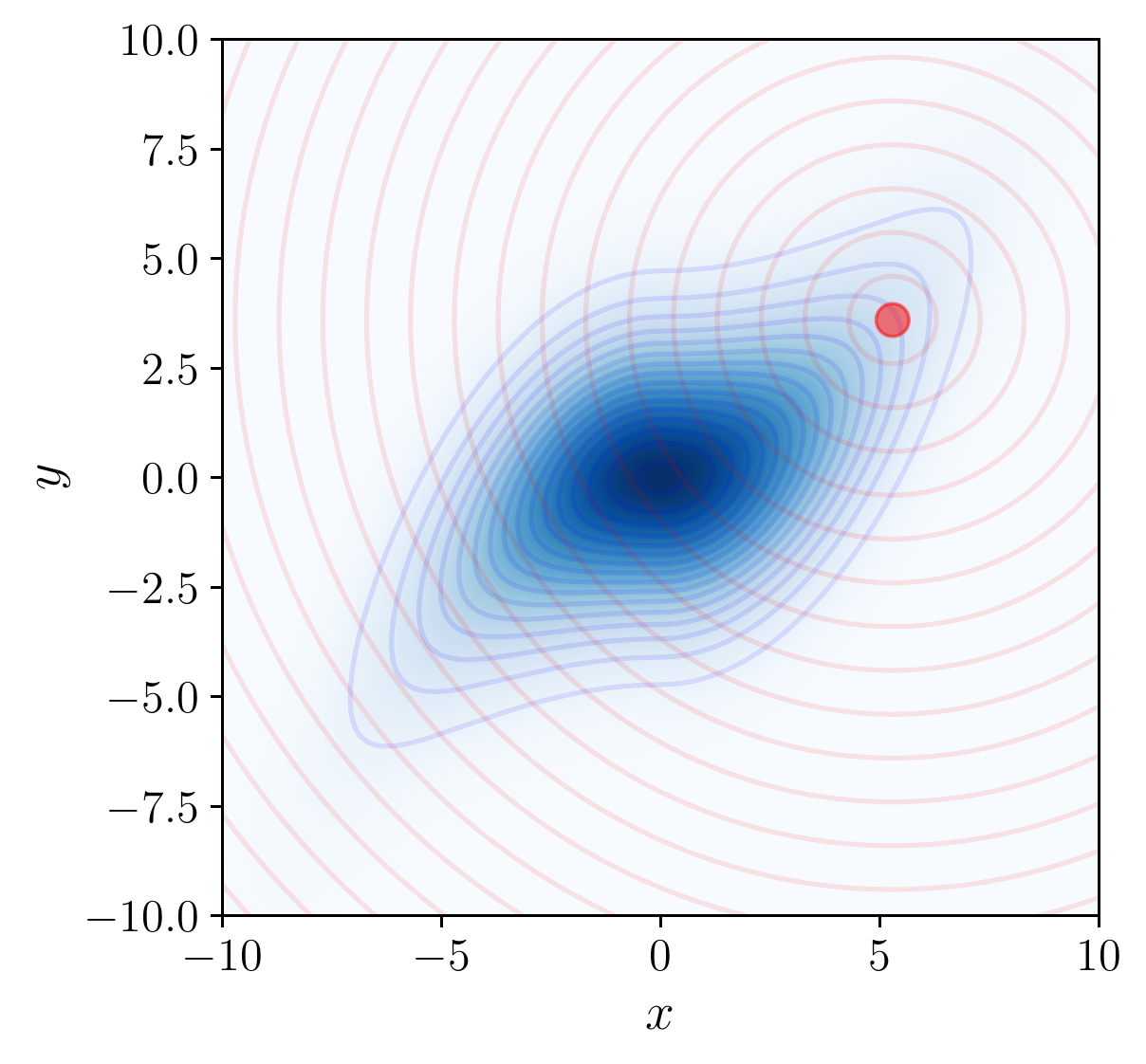}
\caption{To evaluate the log-probability (or density) of the blue
probability distribution {\revision near} the red point, Nested Sampling can be used,
with the blue distribution playing the role of the ``prior'' in NS, and the
Euclidean distance from the red point (illustrated with red contours)
playing the role of the negative log-likelihood. Averaging over selections
of the red point gives an estimate of the entropy of the blue distribution.
\label{fig:algorithm}}
\end{figure}

The idea is to generate a reference particle $\xref$
from the distribution $p(\x)$
whose entropy is required. Then, a Nested Sampling run evolves a set of
particles $\{\x\}_{i=1}^N$, initially representing $p(\x)$,
towards $\xref$ in order to measure the log-probability near
$\xref$ (see Figure~\ref{fig:algorithm}).
Nearness is defined using a distance function
$d(\x; \xref)$, and the number of NS iterations taken to reduce
the distance to below some threshold {\revision $r$ (for `radius')}
provides an unbiased estimate of
\begin{align}
   \depth = -\log \left[ P(d(\x; \xref) < r) \right]
\end{align}
which I call the `depth'. E.g., if $N=10$ Nested Sampling particles are used,
and it takes 100 iterations to reduce the distance to below $r$, then the
depth is 10 nats.

If we actually want the differential entropy
\begin{align}
H(\x) &= -\int p(\x) \log p(\x) \, d\x
\end{align}
we can use the fact that density equals mass divided
by volume. Assume $r$ is small, so that
\begin{align}
P(d(\x; \xref) < r \,|\, \xref)
    &\approx
    p(\xref) \int_{d(\x; \xref) < r} \, d\x\\
    &= p(\xref) \, V
\end{align}
where $V$ is the volume of the region where $d(\x; \xref) < r$.
Then $H(\x) = \left< \depth \right>_{p(\x)} + \log V(r)$.
{\revision It is important to remember that the definition of
volume here is in terms of the coordinates $\x$, and that a change
of coordinates (or a change of the definition of the distance function)
would result in a different value of the
differential entropy. This is equivalent to counting possibilities
or volumes with a different choice of base measure $q(\x)$.}

{\revision Reducing $r$ towards zero ought to lead to
a more accurate result because the probability density at a point will
be more accurately approximated by the probability of a small region containing
the point, divided by the volume of that region.
On the other hand, smaller values of $r$ will lead to more Nested Sampling
iterations required, and hence less accuracy in the depth estimates.
Thus, there is a an awkward trade-off involved in choosing the value of $r$,
which will need to be investigated more thoroughly in the future.
However, the algorithm as it currently stands ought to be useful for calculating
the entropies of any low-to-moderate dimensional functions of an
underlying high-dimensional space.}

\begin{algorithm}
\begin{algorithmic}
\State {\em Set the numerical parameters:}
\State $N \in \{1, 2, 3, ... \}$
            \Comment{the number of Nested Sampling particles to use}
\State $r \geq 0$
            \Comment{the tolerance} \\
\hrulefill
\State $\widehat{\mathbf{h}} \leftarrow \{\}$
            \Comment{Initialise an empty list of results}

\While{more iterations desired}
    \State $k \leftarrow 0$
            \Comment{Initialise counter}
    \State Generate $\xref$ from $p(\x)$
            \Comment{Generate a reference point}
\color{green}
    \State Generate $\left\{\x_i\right\}_{i=1}^N$ from $p(\x)$
            \Comment{Generate initial NS particles}
    \State Calculate $d_i \leftarrow d(\x_i; \xref)$ for all $i$
            \Comment{Calculate distance of each particle from the reference
                     point}
    \State $i^* \leftarrow \textnormal{argmin}(\lambda i \rightarrow d_i)$
            \Comment{Find the worst particle (greatest distance)}
    \State $d_{\rm max} \leftarrow d_{i^*}$
            \Comment{Find the greatest distance}
    \While{$d_{\rm max} > r$}
        \State Replace $\x_{i^*}$ with $\x_{\rm new}$ from
               $p\left(\x \,|\, d(\x) < d_{\rm max}\right)$
            \Comment{Replace worst particle}
        \State Calculate $d_{i^*} \leftarrow d(\x_{\rm new}; \xref)$
            \Comment{Calculate distance of new particle from reference point}
        \State $i^* \leftarrow \textnormal{argmin}(\lambda i \rightarrow d_i)$
                \Comment{Find the worst particle (greatest distance)}
        \State $d_{\rm max} \leftarrow d_{i^*}$
                \Comment{Find the greatest distance}
        \State $k \leftarrow k+1$
            \Comment{Increment counter $k$}
    \EndWhile
\color{black}
    \State {\revision  $\widehat{\mathbf{h}} \leftarrow \textnormal{append}\big(\widehat{\mathbf{h}}, k/N\big)$ }
            \Comment{Append latest estimate to results}
\EndWhile
\State $\widehat{h_{\rm final}} \leftarrow \frac{1}{\rm num\_iterations} \sum \widehat{\mathbf{h}}$
            \Comment{Average results}
\end{algorithmic}
\caption{The algorithm which estimates the expected value of the depth:\\
           \quad\quad$-\int p(\xref) \int p(\x)
            \log \left[ P(d(\x; \xref) < r \,|\, \xref) \right]
                        \, d\x \, d\xref$,\\ that is, minus the expected value of the
        log-probability of a small region near $\xref$, which can be converted
        to an estimate of an entropy or differential entropy.
        The part highlighted in green is standard Nested Sampling
        with quasi-prior $p(\x)$ and quasi-likelihood given by minus a
        distance function $d(\x; \xref)$. The final result, $\widehat{h_{\rm final}}$, is an estimate of the expected depth.
        \label{alg:algorithm}}
\end{algorithm}

If the distance function $d(\xref, \x)$ is chosen to be Euclidean,
the constraint $d(\xref, \x) < r$ corresponds to a ball in the space
of possible $\xref$ values.
The volume of such a ball, of radius $r$ in $n$ dimensions, is
{\revision
\begin{align}
V &= \frac{(\pi r^2)^{n/2}}{(\frac{n}{2})!}.
\end{align}}
In one dimension,
if the threshold $r$ is not small, we might think we are calculating the entropy associated
with the question `what is $\theta$, to within a tolerance of $\pm r$?'.
This is not quite correct. See Appendix~\ref{sec:precisional}
for a discussion of this
type of question.

In the following sections I demonstrate the algorithm on three
example problems. For each of the examples, I used 1000 repetitions of NS
(i.e., 1000 reference particles), and used 10 NS particles and 10,000 MCMC
steps per NS iteration.

\section{Example 1: Entropy of a prior for the data}

This example demonstrates how the algorithm may be used to determine the
entropy for a quantity whose distribution can only be sampled, in this case,
a prior distribution over datasets.

Consider the basic statistics problem of inferring a quantity $\mu$ from
100 observations $\boldsymbol{x} = \{x_1, ..., x_{100}\}$ whose
probability distribution (conditional on $\mu$) is
\begin{align}
p(x_i | \mu) &\sim \textnormal{Normal}(\mu, 1).
\end{align}
{\revision
I assumed a Normal$(0, 10^2)$ prior for $\mu$.
Before getting the data, the prior information is described by the joint
distribution
\begin{align}
p(\mu, \boldsymbol{x}) &= p(\mu)p(\boldsymbol{x} | \mu).\label{eqn:joint_prior}
\end{align}
The prior over datasets is given by the marginal distribution
\begin{align}
p(\boldsymbol{x}) &= \int p(\mu, \boldsymbol{x}) \, d\mu.
\end{align}
This is the distribution whose differential entropy I seek in this section.
Its true value is
available analytically
in this case, by noting that the posterior is available in closed form
for any dataset:
\begin{align}
\mu | \boldsymbol{x} &\sim \textnormal{Normal}\left(
                                       \frac{1}{100.01}\sum_{i=1}^{100} x_i,
                                       \left[\frac{10}{\sqrt{10001}}\right]^2
                                       \right).
\end{align}
which enables the calculation of the mutual information and hence
$H(\boldsymbol{x})$.
The true value of $H(\boldsymbol{x})$ is $146.499$ nats.
However, in most situations, the
entropy of a marginal distribution such as this is not available in closed form.

To use the algorithm, a distance function $d(;)$ must be defined, to quantify
how far the NS particles are from the reference particle. To calculate
the entropy of the data, {\revision I applied the algorithm in the joint
space of possible parameters and datasets}, i.e., distribution of reference
particles followed Equation~\ref{eqn:joint_prior}.
For the distance function, which helps the NS particles approach the
reference particle, I used a
simple Euclidean metric in the space of datasets:
\begin{align}
d\left(\x; \xref\right) &= \sqrt{
        \sum_{i=1}^{100}\left(x_i - x_{{\rm ref}, i}\right)^2}.
\end{align}
Since the distance function refers only to the data $\boldsymbol{x}$
and not the parameter $\mu$, the sampling is effectively done in the
space of possible datasets only --- the parameter $\mu$ can be thought of
as merely a
latent variable allowing $p(\boldsymbol{x})$ to be explored conveniently.
}

I ran the algorithm and computed the average depth using a tolerance of
$r=10^{-3}\sqrt{100}$, so that the RMS difference between
points in the two datasets was about $10^{-3}$.
From the algorithm, the estimate of the expected depth was
\begin{align}
\left<\depth_{\boldsymbol{x}}\right> &= 698.25 \pm 0.34.
\end{align}
The uncertainty here is the standard error of the mean, i.e., the
standard deviation of the depth estimates divided by square root of
the number of NS repetitions {\revision (the number of reference particles
whose depth was evaluated)}.

The log-volume of a 100-dimensional ball of radius
$r=10^{-3}\sqrt{100}$ is
$-551.76$. Therefore, the differential entropy is estimated to be
\begin{align}
H(\boldsymbol{x}) &\approx 146.49 \pm 0.34.
\end{align}
which is very close to the true value (perhaps suspiciously close, but this
was just a fluke).

To carry out the sampling for this problem, Metropolis-Hastings moves that
explore the joint prior distribution $p(\mu, \boldsymbol{x})$ had to be
implemented. Four basic kinds of proposals were used: i) those which change
$\mu$ while leaving $\boldsymbol{x}$ fixed; ii) those which change $\mu$ and
shift $\boldsymbol{x}$ correspondingly; iii) those which resample a subset
of the $x$s from $p(\boldsymbol{x}|\mu)$; and iv) those which move a single
$x$ slightly.

\section{Example 2: Measuring the period of an oscillating signal}

In physics and astronomy, it is common to measure an oscillating signal
at a set of times $\{t_1, ..., t_n\}$, in order to infer the amplitude,
period, and phase of the signal.
Here, I demonstrate the algorithm on a toy version
of Bayesian experimental design: at what times should the signal be
observed in order to obtain as much information as possible about the
period? To answer this, we need to be able to calculate the mutual
information between the unknown period and the data.

As the algorithm lets us calculate the entropy of any distribution which can
be sampled by MCMC, there are several options. The mutual information
can be written in various ways, such as the following.
\begin{align}
I(\x; \d) &= H(\x) + H(\d) - H(\x, \d) \\
I(\x; \d) &= H(\d) - H(\d|\x) \\
I(\x; \d) &= H(\x) - H(\x|\d) \\
I(\x; \d) &= \int p(\d) \, D_{\rm KL}\left(p(\x|\d) \,||\, p(\x)\right) \, d\d\\
I(\x; \d) &= D_{\rm KL}\left(p(\x, \d) \,||\, p(\x)p(\d)\right).
\end{align}
In a Bayesian problem where $\x$ is an unknown parameter and $\d$ is data, the
first and second formulas would be costly to compute, because the
high dimensional probability distribution for the dataset would require a
large number of NS iterations to compute $H(\d)$ and $H(\x, \d)$.
The fourth involves the Kullback-Leibler divergence from the prior to the
posterior, averaged over all possible datasets. This is straightforward to
approximate with standard Nested Sampling, but the estimate for a given dataset
may be biased. {\revision It is quite plausible that the bias is negligible
compared to the variance from averaging over many datasets. This is probably
true in many situations}.

However, more significantly, the above method
would not work if we want $I(\x; \d)$ for
a single parameter or a subset of them, rather than for all of the parameters.
{\revision
The prior-to-posterior KL divergence measured by standard NS is the
amount learned about {\em all} parameters, including nuisance parameters}.
Therefore, I adopted the third formula as the best way to compute $I(\x; \d)$.
In the sinusoidal example, the parameter of interest is the period, whereas
the model will contain other parameters as well.
Since the period will have its prior specified explicitly,
$H(\x)$ will be available analytically. I then use the algorithm to
obtain $H(\x | \d)$ as follows:
\begin{align}
H(\x | \d) &= \int p(\d_{\rm ref}) H(\x | \d=\d_{\rm ref}) \, d\d_{\rm ref}.
\end{align}
This is an expected value over datasets, of the entropy of the posterior given
each dataset. I ran the algorithm by generating reference particles from
$p(\x, \d)$, i.e., generating parameter values from the prior, along with
corresponding simulated datasets. Given each simulated dataset, I used NS
to compute an unbiased estimate of $H(\x | \d=\d_{\rm ref})$, by evolving
a set of particles (initially from the posterior) towards the reference particle.
The role of the ``prior'' in the NS is actually taken by the posterior, and the
``likelihood'' was minus a distance function defined in terms of the
parameter(s) of interest.

To generate the initial
NS particles from the posterior, I used standard MCMC (targeting the
posterior), initialised at the $\x$-value that produced the simulated dataset
$\d_{\rm ref}$, which is a perfect sample from the posterior.

\subsection{Assumptions}

I consider two possible observing strategies, both involving
$n=101$ measurements. The `even' strategy has observations at times
\begin{align}
t_i = \frac{i-1}{n-1}
\end{align}
for $i \in \{1, 2, ..., 101\}$, that is, the observation times are
$0, 0.01, 0.02, ..., 0.99, 1$ ---
evenly spaced from $t=0$ to $t=1$, including the endpoints.
The second, `uneven' observing strategy schedules the observations
according to
\begin{align}
t_i = \left(\frac{i - \frac{1}{2}}{n}\right)^3
\end{align}
which schedules observations close together initially, and further apart
as time goes on.

\begin{figure}[!ht]
\centering
\includegraphics[width=0.7\linewidth]{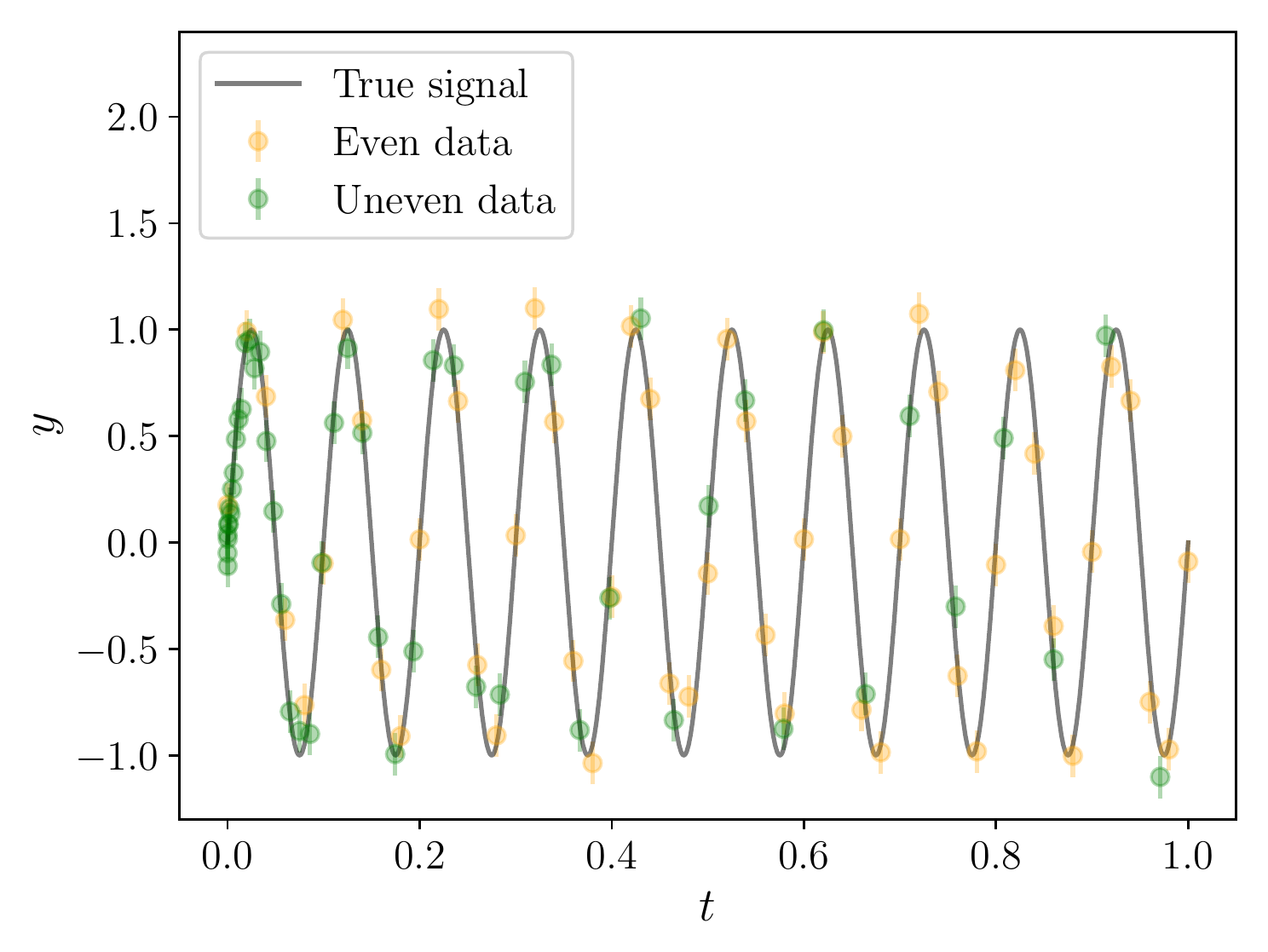}
\caption{A signal with true parameters $A=1$, $\tau=-0.5$, and
$\phi=0$, observed with noise standard deviation 0.1 with the
even (gold points) and uneven (green points) observing strategies.
\label{fig:sinewave}}
\end{figure}

A purely sinusoidal signal has the form
\begin{align}
y(t) &= A \sin \left(\frac{2\pi t}{T} + \phi\right)
\end{align}
where $A$ is the amplitude, $T$ is the period, and $\phi$ is the phase.
Throughout this section, I parameterise the period by its logarithm,
$\tau = \log_{10} T$. I assumed the following priors:
\begin{align}
\ln A   &\sim \textnormal{Normal}\left(0, 0.1^2\right)  \\
\tau    &\sim \textnormal{Uniform}(-1, 0)  \\
\phi    &\sim \textnormal{Uniform}(0, 2\pi)
\end{align}
and the following conditional prior for the data:
\begin{align}
Y_i | A, T, \phi &\sim \textnormal{Normal}\left(y(t_i), 0.1^2\right).
\end{align}
That is, the data is just the signal $y(t)$ observed at particular times
$\{t_1, ..., t_n\}$, with gaussian noise of standard deviation 0.1.
The amplitude of the sinusoid is very likely to be around 10 times the
noise level, and the period is between 0.1 and 1 times the duration of
the data. An example signal observed with the even and uneven observing schedules
is shown in Figure~\ref{fig:sinewave}.

To carry out the sampling for this problem, Metropolis-Hastings moves that
explore the posterior distribution $p(A, \tau, \phi | \boldsymbol{Y})$
had to be implemented. I used heavy-tailed proposals (as in
Section 8.3 of \citet{dnest4}) which modified one parameter at a time.

\subsection{Results}

Letting $\tau = \log_{10} T$, and treating
$A$ and $\phi$ are nuisance parameters, I calculated the conditional entropy
$H(\tau | \boldsymbol{Y})$ using the algorithm with a tolerance of $r = 10^{-5}$.
{\revision The distance function was just the absolute difference between
$\tau$ and $\tau_{\rm ref}$.}

The result was
\begin{align}
\left<\depth\right> &= 5.379 \pm 0.038
\end{align}
Converting to a differential entropy by adding {\revision $\ln (2r) = -10.820$}, the conditional
entropy is
\begin{align}
H(\tau | \boldsymbol{Y}) &= -5.441 \pm 0.038 \textnormal{ nats}.
\end{align}
Since the Uniform(0,1) prior for $\tau$ has a differential entropy of zero, the
mutual information is $I(\tau; \boldsymbol{Y}) = H(\tau) - H(\tau | \boldsymbol{Y}) = 5.441 \pm 0.038$
nats.

The results for the uneven observing schedule were
\begin{align}
\left<\depth\right>       &=  5.422 \pm 0.038 \\
H(\tau | \boldsymbol{Y}) &= -5.398 \pm 0.038 \\
I(\tau; \boldsymbol{Y})   &=  5.398 \pm 0.038 \textnormal{ nats}.
\end{align}

The difference in mutual informations between the two observation
schedules is trivial.
The situation may be different if we
had allowed for shorter periods to be possible, as irregular observing
schedules are known to reduce aliasing and ambiguity in the inferred period
\citep{bretthorst2001nonuniform}.
When inferring the period of an oscillating signal, multimodal posterior
distributions for the period are common \citep{gregoryTrimodal, exoplanet}.
The multimodality of the posterior here raises an interesting issue. Is
the question we really want answered ``what is the value of $T$
{\bf precisely}?'', to which the mutual information relates?
Most practicing scientists would not feel particularly informed to learn
that the vast majority of possibilities had been ruled out, if the
posterior still consisted of several widely separated modes!
Perhaps, in some applications, a more appropriate question is
``what is the value of $T$ to within $\pm$ 10\%'', or something along these
lines. See Appendix~\ref{sec:precisional} for more about this
issue.

In a serious experimental design situation, the relevance is not the only
consideration (or the answer to every experimental design problem would
be to obtain more data without limit), but it is a very important one.

\section{Example 3: Data with Pareto distribution}

Consider a situation where the probability distribution for some data values
$\boldsymbol{x} = \{x_1, x_2, ..., x_n\}$ is a Pareto distribution, conditional on
a minimum value $x_{\rm min}=1$ and a slope $\alpha$:
\begin{align}
p(\boldsymbol{x} | \alpha) &= \prod_{i=1}^n \frac{\alpha}{x_i^{\alpha + 1}}
\end{align}
where all of the $x_i$ are greater than $x_{\rm min} = 1$.
If $\alpha$ is relatively low (close to zero) then this distribution is
very heavy-tailed, and as $\alpha$ increases the probability density
for the $x$s concentrates just above $x=1$.

With heavy-tailed distributions, it can be difficult to predict future
observations from past ones, a notion popularised by
\citet{taleb2007black}. This occurs for three main reasons.
First, the observations may not
provide much information about the parameters (in this case, $\alpha$).
Second, even conditional on the parameters, the predictive distribution for
future data can be heavy-tailed, implying a lot of uncertainty.
Finally, one may always doubt the probability assignment on which such a
calculation is based.
However, the data clearly provides {\em some} nonzero
amount of information about the future.

Consider $n=100$ data values, partitioned into two halves, so
$\boldsymbol{y} = \{x_1, ..., x_{50}\} = \{y_1, ..., y_{50}\}$
is the first half of the data and
$\boldsymbol{z} = \{x_{51}, ..., x_{100}\} = \{z_1, ..., z_{50}\}$
is the second half. Let $\ytot$ and $\ztot$ be the totals of the two
subsets of the data:
\begin{align}
\ytot &= \sum_{i=1}^{50} y_i \\
\ztot &= \sum_{i=1}^{50} z_i.
\end{align}
In this section I use the algorithm to answer the questions:
i) How uncertain is $\ytot$?; ii) How uncertain is $\ztot$?; and
iii) How informative is $\ytot$ about $\ztot$? These are quantified by
$H(\ytot)$, $H(\ztot)$, and $I(\ytot; \ztot)$ respectively. This can
be done despite the fact that $\alpha$ is unknown.

\begin{figure}[!ht]
\centering
\includegraphics[width=0.6\textwidth]{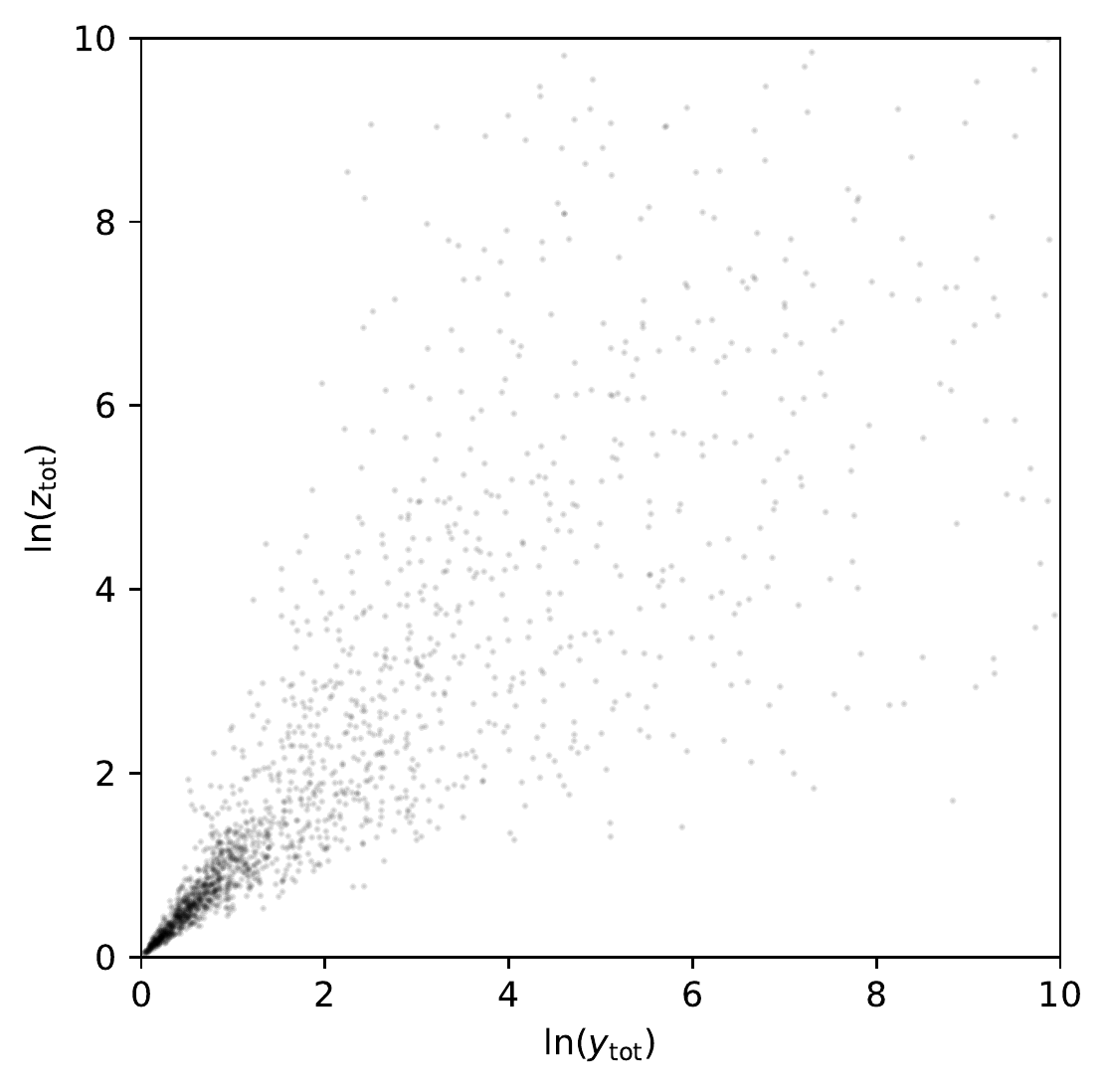}
\caption{The joint distribution for the log-sum of the first half of the
`Pareto data' and the second half. The goal is to calculate the entropy of
the marginal distributions, the entropy of the joint distribution, and
hence the mutual information. The distribution is quite heavy tailed despite
the logarithms, and extends beyond the domain shown in this plot.
\label{fig:pareto}}
\end{figure}

The prior for $\alpha$ was chosen to be lognormal such that the prior median
of $\alpha$ is 1 and the prior standard deviation of $\ln \alpha$ is also 1.
Due to the heavy-tailed nature of the Pareto distribution, I ran the
algorithm in terms of $\ln \ytot$ and $\ln \ztot$. This affects the
differential entropies but not the mutual information (which is invariant 
under changes of parameterisation).
The joint distribution
for $\ln \ytot$ and $\ln \ztot$, whose properties I aimed to obtain,
is shown in Figure~\ref{fig:pareto}.

Three types of Metropolis-Hastings proposals were used for this example:
i) A proposal which changes $\alpha$ and modifies all of the $x$s accordingly;
ii) A proposal which resamples a subset of the $x$s from the Pareto
distribution; and iii) A proposal which moves a single $x$-value slightly.
A fourth proposal which moves $\alpha$ while keeping $\boldsymbol{x}$ fixed,
i.e., a posterior-sampling move, was not implemented.

To compute the entropy of $\ln \ytot$, the distance function was just the
absolute value of the distance between the NS particles and the
reference particle in terms of $\ln \ytot$. The result, using a tolerance
of $r = 10^{-3}$, was:
\begin{align}
\left<\depth_{\ln \ytot}\right> &= 8.851 \pm 0.057.
\end{align}
Correcting for the volume factor yields a differential entropy estimate
of $H(\ln\ytot) = 2.366\pm 0.057$ nats.
By the symmetry of the problem specification, $H(\ln\ztot) = H(\ln\ytot)$.

The joint entropy $H(\ln \ytot, \ln \ztot)$ was also estimated by using
a Euclidean distance function and a tolerance of $10^{-3}$. The result was
\begin{align}
\left<\depth_{(\ln \ytot, \ln \ztot)}\right> &= 16.03 \pm 0.11.
\end{align}
Applying the log-volume correction yields a differential entropy estimate of
$H(\ln \ytot, \ln \ztot) = 3.40 \pm 0.11$ nats.
Therefore, the mutual information is
\begin{align}
I(\ytot; \ztot) &= H(\ln\ytot) + H(\ln\ztot) - H(\ln\ytot, \ln\ztot)\\
                &= 1.130 \pm 0.082 \textnormal{ nats}.
\end{align}
The variance of this mutual information estimate was reduced by
using a common sequence of reference points for the marginal and joint entropy
runs. This is an instance of the `common random numbers'
technique\footnote{\tt http://demonstrations.wolfram.com/TheMethodOfCommonRandomNumbersAnExample/}.
{\revision In short, for those reference particles with a large depth
in terms of the $\ln \ytot$ distance function, the depth in terms of the
$(\ln \ytot, \ln \ztot)$ distance function was also large. Since the
mutual information involves a difference of the averages of these depths,
it is more accurate to compute the average difference 
(i.e. compute the two depths using the same reference particle, to compute
the difference between the two depths) than the difference of the averages.
}

\section{Computational cost, limitations, and conclusion}

The computational resources needed to compute these quantities was quite large,
as Nested Sampling appears in the inner loop of the algorithm.
However, it is worth reflecting on the complexity of the calculation that
has been done.

In a Bayesian inference problem with parameters $\x$ and data $\d$, the
mutual information is
\begin{align}
I(\x; \d) &= \iint p(\x, \d)
                        \ln \left[\frac{p(\x, \d)}{p(\x)p(\d)}\right]
                        \, d\x \, d\d \label{eqn:mutual_info2}
\end{align}
and this measures the dependence between $\x$ and $\d$. A
Monte Carlo
strategy to evaluate this is to sample from $p(\x, \d)$ and average
the value of the logarithm, which involves computing the log-evidence
\begin{align}
\ln p(\d) = \ln\left[\int p(\x)p(\d | \x) \, d\x\right]
\end{align}
for each possible data set.
Computing $p(\d)$ for even a single
data set has historically been considered difficult.

However, if we want the mutual information between the data and
{\em a subset of the parameters} (i.e., there are nuisance parameters
we don't care about), things become even more tricky. Suppose we
partition the parameters $\x$ into important parameters
$\bphi$ and nuisance parameters $\boldeta$, such that we want to calculate
the mutual information $I(\bphi; \d)$. This is given by
\begin{align}
I(\bphi; \d) &= \iint p(\bphi, \d)
                        \ln \left[\frac{p(\bphi, \d)}{p(\bphi)p(\d)}\right]
                        \, d\bphi \, d\d \label{eqn:mutual_info3}
\end{align}
which is equivalent to Equation~\ref{eqn:mutual_info2} but with
$\x$ replaced throughout by $\bphi$. However, this is an even more
difficult calculation than before, as $p(\d | \bphi)$, the marginal likelihood
function with the nuisance parameters integrated out, is typically
unavailable in closed form. If we were to marginalise out the nuisance
parameters $\boldeta$ explicitly, this would give us
Equation~\ref{eqn:mutual_info3} with every probability distribution
written as an explicit integral over $\boldeta$:

\begin{align}
I(\bphi; \d) &= \iint
  \left(\int p(\bphi, \boldeta)p(\d | \bphi, \boldeta) \, d\boldeta\right)
                        \ln \left[
  \frac{\int p(\bphi, \boldeta)p(\d | \bphi, \boldeta) \, d\boldeta}
{p(\bphi)\int p(\bphi, \boldeta)p(\d | \bphi, \boldeta) \, d\bphi \, \d\boldeta}\right]
                        \, d\bphi \, d\d
\end{align}
It should not be surprising that this is costly. Indeed, it seems possible
to me that any {\em general} algorithm for computing entropies will have the
property that NS (or a similar algorithm capable of measuring small volumes)
appears in an inner loop, since entropies are sums of terms, each of which
relates to the downset of a given statement --- the downset of a statement
$S$ being the set of all statements that imply $S$ \citep{knuth_questions}.

{\revision
The current implementation of the algorithm (see Appendix~\ref{sec:software})
is in C++ and has not been parallelised. Given the high computational
costs, parallelisation should be attempted. This could follow
the method of \citep{HENDERSON201784}, or alternatively, different
threads could work on different reference particles simultaneously.}

The algorithm as presented will tend to become less accurate on high dimensional
problems, as the required depth of the runs will increase correspondingly,
since many NS iterations will be required to get a high dimensional vector
of quantities to be close to the reference vector.
The accuracy of the $\ln\depth$ estimates should degrade in proportion to
$\sqrt{\depth}$, due to the Poisson-process nature of the $-\ln X$ sequence.

Throughout this paper, the question of how to sample from the constrained
distributions required by NS has not been discussed in detail. For the examples,
I used the original \citet{skilling2006nested} suggestion of using a short MCMC run
initialised at a surviving point. In some problems, this will not be effective,
and this will cause the resulting estimates to be incorrect, as in standard
Nested Sampling applications. Testing the sensitivity of the results to
changes in the number of MCMC steps used is a simple way to increase one's
confidence in the results, but is not a guarantee {\revision (no general guarantee can
exist, because any algorithm based on sampling a subset of a space might miss
a crucial but hard-to-locate dominant spike)}.
Despite this weakness, I hope the algorithm will be
useful in a wide range of applications.

\acknowledgments{Acknowledgements}
{\revision I would like to thank John Skilling
(Maximum Entropy Data Consultants) and the
anonymous referee for their comments.}
It is a pleasure to thank the following people for interesting and helpful
conversations about this topic, and/or comments on a draft:
Ruth Angus (Flatiron Institute),
Ewan Cameron (Oxford), James Curran (Auckland), Tom Elliott (Auckland),
David Hogg (NYU), Kevin Knuth (SUNY Albany),
Thomas Lumley (Auckland),
Iain Murray (Edinburgh), Jared Tobin ({\tt jtobin.io}).
This work was supported by Centre for eResearch
at the University of Auckland.




\conflictofinterests{Conflicts of Interest}
The authors declare no conflicts of interest.

\makeatletter
\renewcommand\@biblabel[1]{#1. }
\makeatother

\bibliographystyle{mdpi}

\begin{thebibliography}{-------}
\providecommand{\natexlab}[1]{#1}

\bibitem[Shannon(1948)]{shannon}
Shannon, C.E.
\newblock A mathematical theory of communication.
\newblock {\em The Bell System Technical Journal} {\bf 1948}, {\em
  27},~379--423.

\bibitem[Cover and Thomas(2012)]{cover2012elements}
Cover, T.M.; Thomas, J.A.
\newblock {\em Elements of information theory}; John Wiley \& Sons,  2012.

\bibitem[Knuth(2005)]{knuth_questions}
Knuth, K.H.
\newblock Toward question-asking machines: the logic of questions and the
  inquiry calculus {\bf 2005}.

\bibitem[Knuth and Skilling(2012)]{knuth2012foundations}
Knuth, K.H.; Skilling, J.
\newblock Foundations of inference.
\newblock {\em Axioms} {\bf 2012}, {\em 1},~38--73.

\bibitem[Caticha and Giffin(2006)]{caticha2006updating}
Caticha, A.; Giffin, A.
\newblock Updating Probabilities.
\newblock  Bayesian Inference and Maximum Entropy Methods In Science and
  Engineering. AIP Publishing,  2006, Vol. 872, pp. 31--42.

\bibitem[Szab\'{o}(2014)]{JMLR:v15:szabo14a}
Szab\'{o}, Z.
\newblock Information Theoretical Estimators Toolbox.
\newblock {\em Journal of Machine Learning Research} {\bf 2014}, {\em
  15},~283--287.

\bibitem[Jaynes(2003)]{jaynes2003probability}
Jaynes, E.T.
\newblock {\em Probability theory: The logic of science}; Cambridge university
  press,  2003.

\bibitem[MacKay(2003)]{mackay2003information}
MacKay, D.J.
\newblock {\em Information theory, inference and learning algorithms};
  Cambridge university press,  2003.

\bibitem[Caticha(2008)]{caticha2008lectures}
Caticha, A.
\newblock Lectures on probability, entropy, and statistical physics.
\newblock {\em arXiv preprint arXiv:0808.0012} {\bf 2008}.

\bibitem[Bernardo(2005)]{bernardo2005reference}
Bernardo, J.M.
\newblock Reference analysis.
\newblock {\em Handbook of statistics} {\bf 2005}, {\em 25},~17--90.

\bibitem[Skilling(2006)]{skilling2006nested}
Skilling, J.
\newblock Nested sampling for general Bayesian computation.
\newblock {\em Bayesian analysis} {\bf 2006}, {\em 1},~833--859.

\bibitem[Feroz \em{et~al.}(2009)Feroz, Hobson, and Bridges]{feroz2009multinest}
Feroz, F.; Hobson, M.; Bridges, M.
\newblock MultiNest: an efficient and robust Bayesian inference tool for
  cosmology and particle physics.
\newblock {\em \mnras} {\bf 2009}, {\em 398},~1601--1614.

\bibitem[Brewer \em{et~al.}(2011)Brewer, Pártay, and Csányi]{dns}
Brewer, B.J.; Pártay, L.B.; Csányi, G.
\newblock Diffusive nested sampling.
\newblock {\em Statistics and Computing} {\bf 2011}, {\em 21},~649--656.

\bibitem[Handley \em{et~al.}(2015)Handley, Hobson, and
  Lasenby]{handley2015polychord}
Handley, W.; Hobson, M.; Lasenby, A.
\newblock POLYCHORD: next-generation nested sampling.
\newblock {\em \mnras} {\bf 2015}, {\em 453},~4384--4398.

\bibitem[Knuth \em{et~al.}(2015)Knuth, Habeck, Malakar, Mubeen, and
  Placek]{knuth2015bayesian}
Knuth, K.H.; Habeck, M.; Malakar, N.K.; Mubeen, A.M.; Placek, B.
\newblock Bayesian evidence and model selection.
\newblock {\em Digital Signal Processing} {\bf 2015}, {\em 47},~50--67.

\bibitem[Pullen and Morris(2014)]{pullen2014bayesian}
Pullen, N.; Morris, R.J.
\newblock Bayesian model comparison and parameter inference in systems biology
  using nested sampling.
\newblock {\em PloS one} {\bf 2014}, {\em 9},~e88419.

\bibitem[Brewer and Donovan(2015)]{exoplanet}
Brewer, B.J.; Donovan, C.P.
\newblock Fast Bayesian inference for exoplanet discovery in radial velocity
  data.
\newblock {\em \mnras} {\bf 2015}, {\em 448},~3206--3214.

\bibitem[Pártay \em{et~al.}(2010)Pártay, Bartók, and
  Csányi]{partay2010efficient}
Pártay, L.B.; Bartók, A.P.; Csányi, G.
\newblock Efficient sampling of atomic configurational spaces.
\newblock {\em The Journal of Physical Chemistry B} {\bf 2010}, {\em
  114},~10502--10512.

\bibitem[Baldock \em{et~al.}(2016)Baldock, Pártay, Bartók, Payne, and
  Csányi]{baldock2016determining}
Baldock, R.J.; Pártay, L.B.; Bartók, A.P.; Payne, M.C.; Csányi, G.
\newblock Determining pressure-temperature phase diagrams of materials.
\newblock {\em Physical Review B} {\bf 2016}, {\em 93},~174108.

\bibitem[Martiniani \em{et~al.}(2014)Martiniani, Stevenson, Wales, and
  Frenkel]{martiniani2014superposition}
Martiniani, S.; Stevenson, J.D.; Wales, D.J.; Frenkel, D.
\newblock Superposition enhanced nested sampling.
\newblock {\em Physical Review X} {\bf 2014}, {\em 4},~031034.

\bibitem[Henderson \em{et~al.}(2017)Henderson, Goggans, and
  Cao]{HENDERSON201784}
Henderson, R.W.; Goggans, P.M.; Cao, L.
\newblock Combined-chain nested sampling for efficient Bayesian model
  comparison.
\newblock {\em Digital Signal Processing} {\bf 2017}, {\em 70},~84 -- 93.

\bibitem[Walter(2017)]{Walter2017point}
Walter, C.
\newblock Point process-based Monte Carlo estimation.
\newblock {\em Statistics and Computing} {\bf 2017}, {\em 27},~219--236.

\bibitem[{Brewer} and {Foreman-Mackey}(2016)]{dnest4}
{Brewer}, B.J.; {Foreman-Mackey}, D.
\newblock {DNest4: Diffusive Nested Sampling in C++ and Python}.
\newblock {\em Journal of Statistical Software, accepted. arxiv: 1606.03757}
  {\bf 2016},  \href{http://xxx.lanl.gov/abs/1606.03757}{{\normalfont
  [arXiv:stat.CO/1606.03757]}}.

\bibitem[Bretthorst(2001)]{bretthorst2001nonuniform}
Bretthorst, G.L.
\newblock Nonuniform sampling: Bandwidth and aliasing.
\newblock  AIP conference proceedings. AIP,  2001, Vol. 567, pp. 1--28.

\bibitem[{Gregory}(2005)]{gregoryTrimodal}
{Gregory}, P.C.
\newblock {A Bayesian Analysis of Extrasolar Planet Data for HD 73526}.
\newblock {\em \apj} {\bf 2005}, {\em 631},~1198--1214.

\bibitem[Taleb(2007)]{taleb2007black}
Taleb, N.N.
\newblock The Black Swan: The Impact of the Highly Improbable.
\newblock {\em ISBN: 978-1400063512} {\bf 2007}.

\end{thebibliography}

\appendix
\section{Precisional questions}
\label{sec:precisional}

The central issue about the value of a parameter $\x$ asks
``what is the value of $\x$, precisely?''. However, in practice we often
don't need or want to know $\x$ to arbitrary precision. Using the central
issue can lead to counterintuitive results if you don't keep its specific
definition in mind. For example, suppose $x$ could take any integer value
from 1 to 1 billion. If we learned the last digit of $x$, we
will have ruled out nine tenths of the possibilities, and therefore obtained
a lot of information about the central issue. However,
this information might be useless {\em for practical purposes}.
If we learned the final
digit was a 9, $x$ could still be 9, or 19, or 29, any number ending in 9
up to 999,999,999.

In practice, we may
want to ask a question that is different from the central issue.
For example, suppose $x \in \{1, ..., 10\}$ and
we want to know the value of $x$ to within a tolerance
of $\pm 1$. Any of the following statements would
resolve the issue:
\begin{itemize}
\item $x \in \{1, 2, 3\}$ and anything that implies it
\item $x \in \{2, 3, 4\}$ and anything that implies it
\item $x \in \{3, 4, 5\}$ and anything that implies it
\item and so on.
\end{itemize}
According to \citet{knuth_questions}, the entropy of a question is
computed by applying the sum rule over all statements that would answer the
question.

We can write the question of interest, $Q$, as a union of ideal questions
(the ideal questions are downsets of statements: the set of all statements
that imply $S$ is the downset of $S$ and is denoted $\downarrow S$):
\begin{align}
Q &= \left[\downarrow (x \in \{1, 2, 3\})\right] \cup
     \left[\downarrow (x \in \{2, 3, 4\})\right] \cup
     \left[\downarrow (x \in \{3, 4, 5\})\right] \cup ...
\end{align}
The entropy of this question is
\begin{align}
H(Q) &= h_{1,2,3}\\
     & \quad \quad + \left(h_{2,3,4} - h_{2,3}\right) \\
     & \quad \quad + \left(h_{3,4,5} - h_{3,4}\right) \\
     & \quad \quad + ... \\
     & \quad \quad + \left(h_{8,9,10} - h_{8,9}\right).
\end{align}
where $h_{x} = -P(x)\log P(x)$.

\subsection*{Continuous case}

Consider a probability density function $f(x)$, defined on the real
line. The probability contained in an interval
$[x_0, x_0 + r]$, which has length $r$, is
\begin{align}
\int_{x_0}^{x_0 + r} f(x) \, dx &= F(x_0 + r) - F(x_0)
\end{align}
where $F(x)$ is the cumulative distribution function (CDF).
This probability can be considered as a function of two variables,
$x_0$ and $r$, which I will denote by $P(,)$:
\begin{align}
P(x_0, r) &= F(x_0 + r) - F(x_0).
\end{align}
The contribution of such an interval to an entropy expression
is $-P\log P$, i.e.,
\begin{align}
Q(x_0, r) &= -P(x_0, r) \log P(x_0, r).
\end{align}
Consider the rate of change of $Q$ as the interval is shifted to
the right, but with its width $r$ held constant:
\begin{align}
\frac{\partial Q}{\partial x_0} &= \frac{\partial}{\partial x_0}
    \left[-P\log P\right] \\
    &= -\left[1 + \log P(x_0, r)\right]\frac{\partial P}{\partial x_0} \\
    &= -\left[1 + \log P(x_0, r)\right]\left[f(x_0 + r) - f(x_0)\right].
\end{align}
Consider also the rate of change of $Q$ as the interval is
expanded to the right, while keeping its left edge fixed:
\begin{align}
\frac{\partial Q}{\partial r} &= \frac{\partial}{\partial r}
    \left[-P(x_0, r) \log P(x_0, r)\right] \\
    &= -\left[1 + \log P(x_0, r)\right]\frac{\partial P}{\partial r} \\
    &= -\left[1 + \log P(x_0, r)\right]f(x_0 + r).
\end{align}

The entropy of the precisional question is built up from
$Q$ terms. The extra entropy from adding the interval
$[x_0, x_0 + r]$ and removing the overlap $[x_0, x_0 + r - h]$, for small $h$, is
\begin{align}
\delta H &= Q(x_0, r) - Q(x_0, r - h)\\
         &= h\frac{\partial Q}{\partial r} \\
         &= -h\left[1 + \log P(x_0, r)\right]f(x_0 + r).
\end{align}
Therefore, the overall entropy is
\begin{align}
H &= -\int_{-\infty}^\infty
        \left[1 + \log P(x-r, r)\right]f(x)
      \, dx \\
  &= -\int_{-\infty}^\infty
        \left(1 + \log\left[F(x) - F(x-r)\right]\right)f(x)
      \, dx.
\end{align}
Interestingly, calculating the log-probability of an interval
to the right of $x$ gives an equivalent result:
\begin{align}
H' &=  -\int_{-\infty}^\infty
        \left(1 + \log\left[F(x+r) - F(x)\right]\right)f(x)
      \, dx \\
   &= H.
\end{align}
This is not quite the same as the expected log-probability calculated by
the version of the algorithm proposed in this paper (when the tolerance is not
small). However, the algorithm can be made to estimate the entropy of the
precisional question, by redefining the distance function. {\revision For a one-dimensional
quantity of interest, the distance function can be defined such that
$d(x; x_{\rm ref})$ can only ever be below $r$ if $x < x_{\rm ref}$ (or alternatively,
if $x > x_{\rm ref}$).}

\section{Software}
\label{sec:software}

A {\tt C++} implementation of the algorithm is available in a {\tt git}
repository located at
\begin{verbatim}
https://github.com/eggplantbren/InfoNest
\end{verbatim}
and can be obtained using the following {\tt git} command, executed in a
terminal:
\begin{verbatim}
git clone https://github.com/eggplantbren/InfoNest
\end{verbatim}
The following will compile the code and execute the first example from the
paper:
\begin{verbatim}
cd InfoNest/cpp
make
./main
\end{verbatim}
The algorithm will run for 1000 `reps', i.e., 1000 samples of $\xref$, which is
time consuming. Output is saved to {\tt output.txt}. At any time,
you can execute the Python script {\tt postprocess.py} to get an estimate of
the depth:
\begin{verbatim}
python postprocess.py
\end{verbatim}

By default, {\tt postprocess.py} estimates the depth with a tolerance of
$r=10^{-3}$. This value can be changed by calling the {\tt postprocess}
function with a different value of its argument {\tt tol}. E.g.,
{\tt postprocess.py} can be edited so that its last line is
{\tt postprocess(tol=0.01)} instead of {\tt postprocess()}.

The numerical parameters, and the choice of which problem is being solved,
are specified in {\tt main.cpp}. The default problem is the first example
from the paper. For Example 2, since it is a conditional entropy which is
being estimated (which requires a slight modification to the algorithm),
an additional argument {\tt InfoNest::Mode::conditional\_entropy}
must be passed to the {\tt InfoNest::execute} function.


%


\end{document}